\documentclass[epj]{webofc}
\usepackage[varg]{txfonts}   
\usepackage{graphicx,color} 
\usepackage{bm}       
\usepackage{amsmath}  
\usepackage{amssymb}
\woctitle{21st International Conference on Few-Body Problems in Physics}
\begin{document}
\begin{flushright}
YITP-16-25
\end{flushright}
\title{Hadron Interactions from lattice QCD}
\author{Sinya Aoki\inst{1,2}\fnsep\thanks{\email{saoki@yukawa.kyoto-u.ac.jp}} 
}

\institute{Yukawa Institute for Theoretical Physics, Kyoto University, Kyoto 606-8502, Japan
\and
           Center for Computational Sciences, University of Tsukuba. Ibaraki 305-8571, Japan
          }

\abstract{
  We review our strategy to study hadron interactions from lattice QCD using newly proposed potential method. 
  We first explain our strategy in the case of nuclear potentials and its application to nuclear physics. We then discuss the origin of the repulsive core, by adding strange quarks to the system. We also explore a possibility for H-dibaryon to exist in flavor SU(3) limit of lattice QCD. We conclude the paper with an application of our strategy to investigate the maximum mass of neutron stars.
}
\maketitle
\section{Introduction}
\label{intro}
The current understanding of the structure of matters is as follows.
At some microscopic level, the building blocks of matters are  atoms, which are composed of heavy and compact nuclei in the center surrounded by electrons. A nucleus is made of protons and neutrons, which are also composed of more fundamental objects, quarks. A word hadron is a generic name for composite particles made of quarks such as proton and neutron.

\begin{figure}[hbt]
\centering
\sidecaption
\includegraphics[width=5cm,clip]{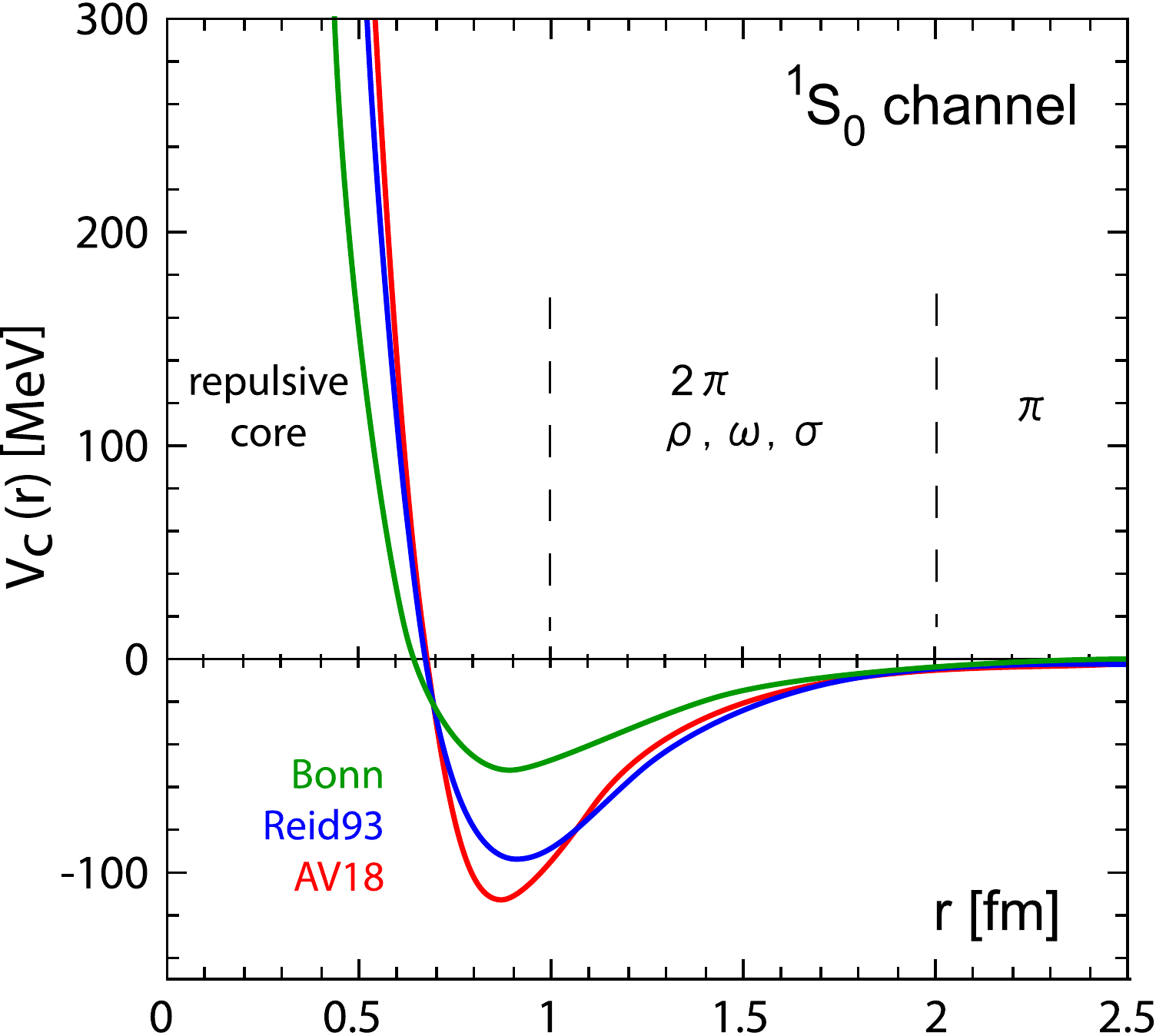}
\caption{Three examples of  the  phenomenological nuclear potential,  Bonn\cite{Machleidt:2000ge}, Reid93\cite{Stoks:1994wp} and Argonne $v_{18}$\cite{Wiringa:1994wb}.  Taken from Ref.~\cite{Ishii:2006ec}. }
\label{fig-1}
\vskip -0.3cm       
\end{figure}

In this talk, we consider interactions among such hadrons. For example, protons and neutrons interact with each other to form nuclei, bound states of them. Such interactions is called the nuclear force, the origin of which 
has been explained by the exchange of new particles, called pion\cite{Yukawa1935}.
We now know not only proton and neutron, collectively called nucleons,  but also pions are composite particles of quarks, together with gluons, which mediate the strong force between quarks. 
Fig.~\ref{fig-1} shows some phenomenological estimates of the nuclear potential as a function of a distance between two nucleons, whose derivative with respect to the distance gives the force between two nucleons, the nuclear force.
As can be seen from the figure, the potential shows a rather complicated behavior as a function of the distance due to the compositeness of nucleons. 

The dynamics of quarks and gluons is known to be governed by the theory called Quantum Chromodynamics (QCD).  Therefore, in principle, hadron interactions such as the nuclear force can be derived from QCD.
Interactions of QCD becomes so strong at long distance that no isolated quarks can be observed. This phenomenon is called  quark confinement.  The perturbative expansion relied on the weak coupling  does not work for QCD since
its coupling constant is large as the quark confinement indicates.
Lattice field theories have been proposed to investigate strongly interacting theories such as QCD, by defining quantum field theories on a discrete space-time (lattice) instead of the continuous one.  
QCD defined on $N^4$ lattice (lattice QCD) is equivalent to well-defined statistical system with a finite lattice spacing $a$ and the finite extension $L= N a$, which is manifestly gauge invariant and non-perturbatively defined\cite{Wilson:1974sk}. Because of these properties, numerical simulations based on Monte-Carlo method can be successfully applied to lattice QCD\cite{Creutz:1980zw}. 

Fundamental degrees of freedom in QCD are quarks and gluons, from which stable hadrons such as nucleon and pions emerge as physical degrees of freedom by QCD interactions.
Interactions among these stable particles  are also described by the same theory, QCD.
A main purpose of this talk is to review our attempts to understand hadron interactions based on QCD by using lattice QCD techniques.

\section{Strategy and nuclear potentials}
\label{sec-2}

\subsection{Our strategy}
\label{sec-2-1}
We first briefly explain our strategy to investigate hadron interactions in lattice QCD\cite{Ishii:2006ec,Aoki:2009ji}, which
consists of three main steps.

In the 1st step, we calculate potentials between two hadrons by numerical simulations in lattice QCD. 
There are many subtleties and discussions on how to define potentials in QCD, but we do not discuss them in this talk. Please see Ref.~\cite{Aoki:2012tk} for more details.

In the 2nd step, we calculate physical observables such as binding energy  and scattering phase shift of the two hadrons, using potentials obtained in the 1st step.
 
In the 3rd step, we employ the same 2-body potentials to investigate few hadron systems or structures of may body systems such as heavy nuclei and nuclear matters.  

The 1st step in the above  has been first applied to the nuclear potential~\cite{Ishii:2006ec}, which has shown that the proposal seems  working well, and Ref.~\cite{Ishii:2006ec} has received general recognition\cite{nature}.
After this success, a research group named Hadrons to Atomic nuclei from Lattice QCD (HAL QCD) Collaboration has bend formed to investigate various aspects of hadronic interactions by this method, called the HAL QCD method.\footnote{The binding energy of two nucleon systems has been calculated directly in lattice QCD (for example, see \cite{Yamazaki:2012hi,Beane:2011iw}), but results disagree with ours. 
Recently, however, it turns out that plateaux of the binding energy in the finite box from two different source operators,  wall and smeared ones, differ significantly\cite{Iritani:2015},
so that the claimed strong binding  of the two nucleon systems at heavy pion masses \cite{Yamazaki:2012hi,Beane:2011iw} should be taken with a grain of salt. We therefore do not consider the direct method in this report. }
  
\subsection{Example: Nuclear potentials}
\label{sec-2-2}
Let us consider the nuclear potential, as an example how the strategy in Sect.~\ref{sec-2-1} works.

\begin{figure}[bth]
\vskip -0.5cm       
\centering
\sidecaption
\includegraphics[width=6.5cm,clip]{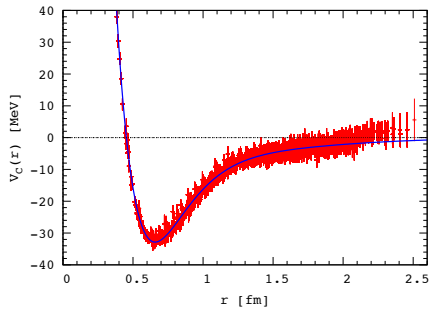}
\caption{Spin-singlet nuclear potential calculated in lattice QCD, together with the multi-Gaussian fit by the solid line. Taken from Ref.~\cite{HALQCD:2012aa}.  }
\label{fig-2}       
\vskip -0.5cm       
\end{figure}

In the 1st step, we calculate the nuclear potential in lattice QCD. Fig.~\ref{fig-2} shows the nuclear potential for the spin-singlet ($S=0$) sector, calculated in lattice QCD at the lattice spacing (space resolution) $a\simeq $ 0.1 fm with $m_u=m_d < m_s$, where $m_u, m_d, m_s$ are up, down, strange quark masses, respectively\cite{HALQCD:2012aa}.  In this calculation, pion mass $m_\pi$, which is a measure of quark mass in the simulation, is about 700 MeV, while $m_\pi \simeq 140$ MeV in Nature.
Lighter the quark mass, harder the simulation in lattice QCD. This is a reason why we employed $m_\pi\simeq 700$ MeV  for this study. 
As can be seen from Fig.~\ref{fig-2}, qualitative features of nuclear potential in Fig.~\ref{fig-1} are reproduced in our lattice QCD simulations. Namely, the potential in Fig.~\ref{fig-2} shows attractions at medium and long distances, while it gives repulsions at short distance, called a repulsive core. 

\begin{figure}[bth]
\vskip -0.3cm       
\centering
\sidecaption
\includegraphics[width=6.5cm,clip]{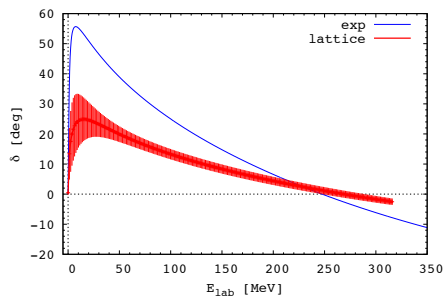}
\caption{The  scattering  phase in  the spin-singlet channel in  the
  laboratory  frame obtained  from the  lattice nuclear  potential in Fig.~\ref{fig-2}, together
  with experimental data by the solid line\cite{nn-online}.  Taken from Ref.~\cite{HALQCD:2012aa}.  }
\label{fig-3}       
\vskip -0.3cm       
\end{figure}

In the 2nd step, we employ the potential obtained in lattice QCD to calculate the scattering phase shift of two nucleons. In practice, the potential in Fig.~\ref{fig-2} is fitted with some functional form. For example, the multi-Gaussian function is used in Fig.~\ref{fig-2}, shown by the solid line. We then solve the Schr\"odinger equation to determine the phase shift at a given energy.  Fig.~\ref{fig-3} shows the scattering phase shift in the spin-singlet channel as a function of the laboratory energy $E_{\rm lab}$ with the rest energy of two nucleons subtracted\cite{HALQCD:2012aa}.
The scattering phase shift obtained form lattice QCD has a reasonable shape. Its strength, however, is weaker than the experimental one (solid line), probably due to the heavier quark mass in lattice QCD simulations:
The scattering length, calculated as $\lim_{k\rightarrow 0} \tan\delta(k) /k$, is  $1.6\pm 1.1$ fm in this lattice QCD simulation at $m_\pi \simeq 700$ MeV, while it should be about 20 fm at the physical pion mass, $m_\pi \simeq 140$ MeV.
Lattice QCD simulations at the physical pion mass  are definitely required to reproduce the correct scattering phase shift and scattering length in lattice QCD.
Such attempts are currently ongoing by using   the K computer in Kobe, Japan, whose peak speed is 10 PetaFlops\cite{K-computer}. 

\begin{figure}[hbt]
\centering
\sidecaption
\includegraphics[width=7cm,clip]{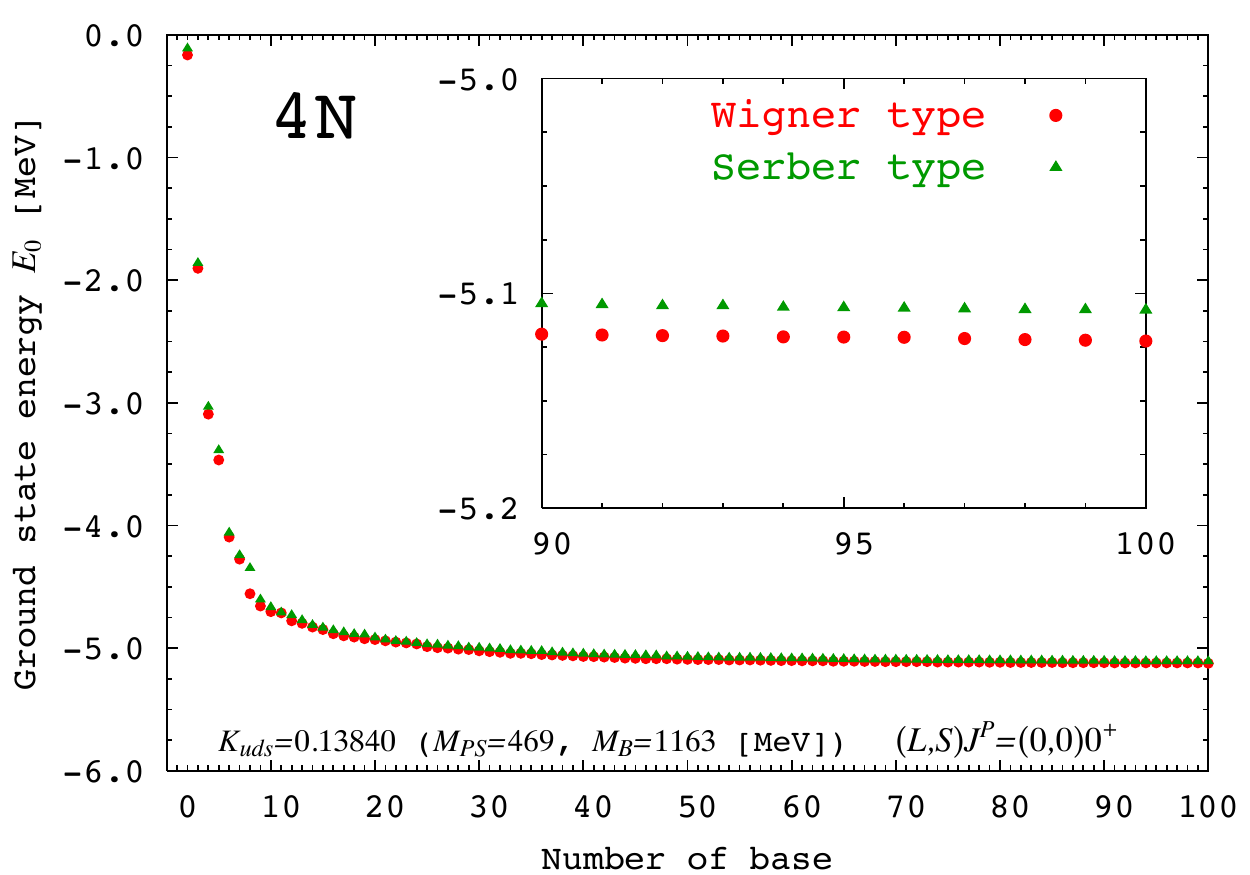}
\caption{Biding energy of 4 nucleon ground state in lattice QCD with the flavor SU(3) limit, as a function of a number of bases. Two types of 2-body potentials are employed for a comparison. Taken from Ref.~\cite{Inoue:2011ai}.  }
\label{fig-4}       
\end{figure}

In the 3rd step, using the potential obtained in lattice QCD, we estimate the binding energy of some nuclei.
For example, Fig.~\ref{fig-4} shows the binding energy of a 4 nucleon ground state in $(L,S)J^P =(0,0)0^+$,
configuration corresponding to the $^4$He nucleus, as a function of a number of bases in the stochastic variational method\cite{Inoue:2011ai}. The potentials used in this calculation, different from the one in Fig.~\ref{fig-2},
are obtained in the flavor SU(3) limit that $m_u=m_d=m_s$ at $a\simeq 0.12$ fm and $L \simeq 4$ fm with
$M_{\rm PS} \simeq 470$ MeV, where $M_{\rm PS}$ is the mass of pseudo-scalar meson ($\pi$ and K). 
In the calculation of the ground state energy for $^4$He, the Wigner type force, where the odd part  of potentials is equal to the even part, as well as the Serber  type force, where the odd part is set to zero, are employed for a comparison. As seen in Fig.~\ref{fig-4}, results form two different forces agree, and the binding energy of $^4$He is about 5.1 MeV.  Again, a smaller binding energy than Nature is partly due to the heaver quark masses and to the flavor SU(3) limit, in addition to absence of 3- and 4- nucleon forces.
See Ref.~\cite{Inoue:2011ai} for more details. 
Recent applications of lattice QCD potentials to  medium-heavy nuclei can be found in Ref.~\cite{Inoue:2014ipa}.

\section{Repulsive core}
The short distance repulsion in the nuclear force, first introduced by Jastrow\cite{jastrow} and called the repulsive core,  plays several important roles in nuclear physics and astrophysics. 
For example, it explains stability of matters.  A nucleus would collapse to a very dense object if the nuclear force had only attractions without repulsive core. 
The repulsive core sustains neutron stars against their gravitational collapses and
ignites   the Type II supernova explosions as the gravitational collapse bounds by the repulsion\cite{VJ}.

\subsection{The origin of the repulsive core}
\begin{figure}[hbt]
\centering
\sidecaption
\includegraphics[width=7cm,clip]{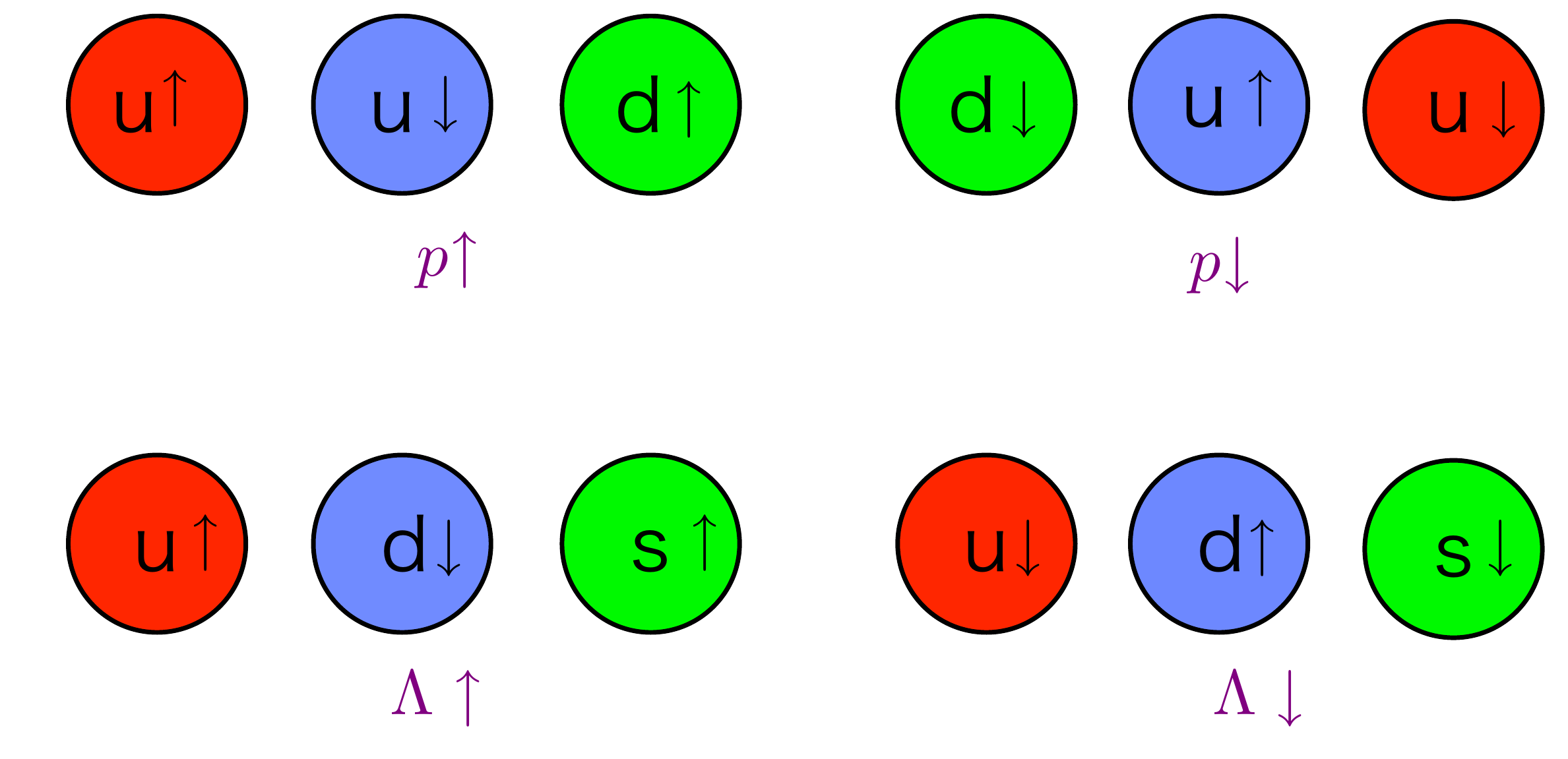}
\caption{(Top) An example of quark configurations for two protons with opposite spins.
(Bottom) An example of quark configurations for two $\Lambda$'s with opposite spins.
}
\label{fig-5}       
\end{figure}

It is interesting and important to ask what is the origin of the repulsive core.
Since quarks are fermions, two can not occupy the same position with same quantum numbers due to the Pauli exclusion principle.
However, quarks have 3 colors (red, blue and green),  2 spins ($\uparrow$ and $\downarrow$) and 2 flavors (up and down), so that  six quarks can occupy the same position. Since this means that two nucleons occupy the same position, the repulsive core can not be explained simply by the Pauli exclusion principle alone. 
However, allowed color combinations are restricted once spin and flavor quantum numbers are fixed, as an example for two protons with opposite spins is shown in the top of Fig.~\ref{fig-5}. 
In this example, up quarks with $\uparrow$  in each proton can not have the same color.
Similarly, up quarks  with $\downarrow$ also can not. 
It has been pointed out that this restriction for color combinations plus one gluon-exchange interactions between quarks may be an origin of the repulsive core\cite{Oka:1983ku, Oka:1986fr}.

In order to check whether the origin of the repulsive core is related to the partial restriction of color combinations, it is useful to consider what happen if an extra degree of freedom, a strange quark, is added to the system. In this system, we have the $\Lambda(uds)$ baryon, which is composed of up, down and strange quarks. If we consider two $\Lambda$'s with opposite spins, all color combinations are allowed as drawn in the bottom of  Fig.~\ref{fig-5}, where two  quarks with the same flavor (up, down or strange) in each $\Lambda$ have same color. If the origin of the repulsive core is related to the partial restriction of color combinations,
the potential between two $\Lambda$'s in the spin singlet channel has no repulsive core.

\begin{figure}[hbt]
\centering
\sidecaption
\includegraphics[width=7cm,clip]{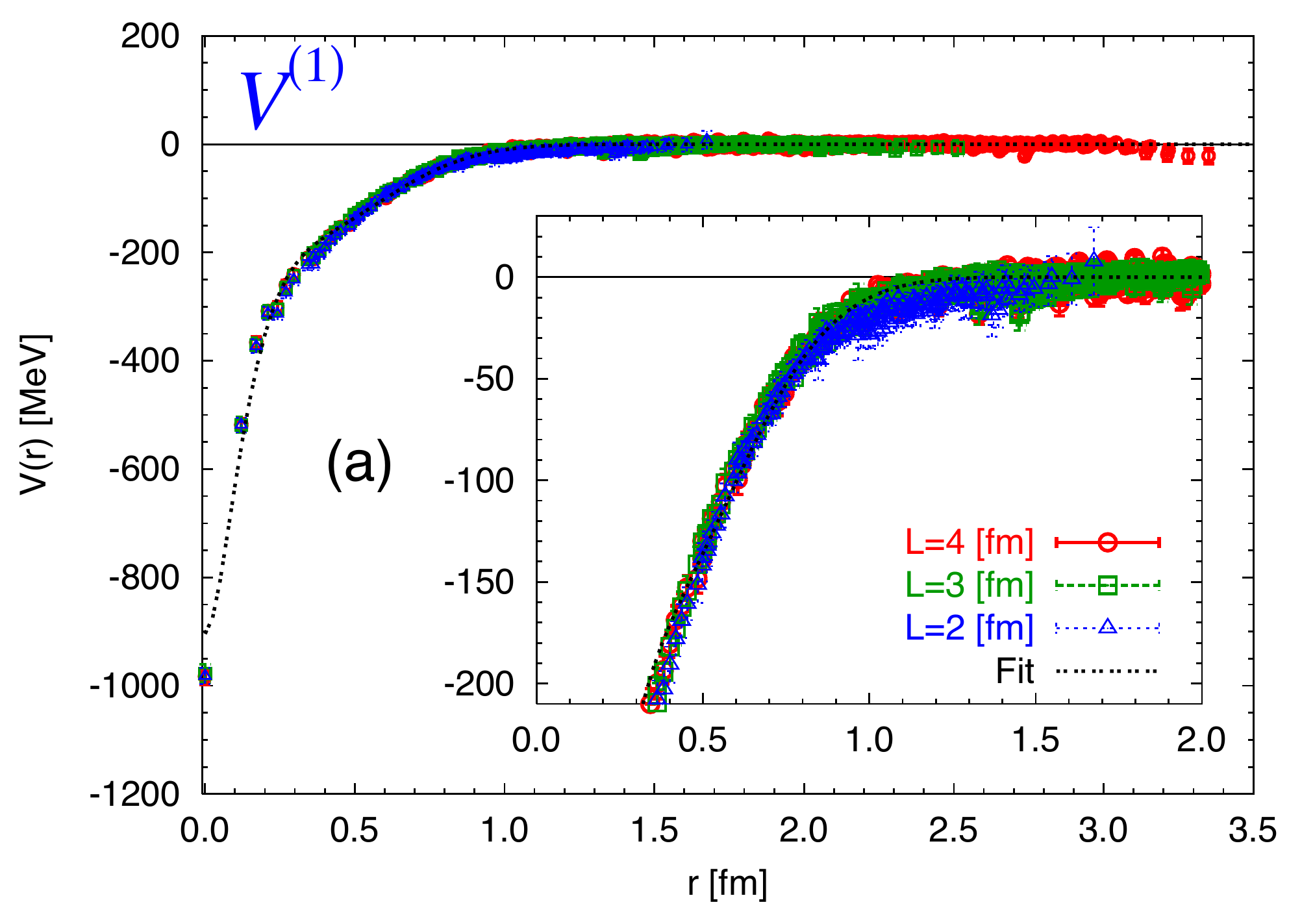}
\caption{Flavor-singlet potential $V_C^{(1)}(r)$ for the box size $L\simeq$ 2, 3, 4 fm at $M_{\rm PS}=$ 1015 MeV. Taken from Ref.~\cite{Inoue:2010es}.  }
\label{fig-6}       
\end{figure}
To confirm this, we have performed 3-flavor lattice QCD simulations in the flavor SU(3) limit ($m_u=m_d=m_s$)\cite{Inoue:2010hs,Inoue:2010es}.
In the flavor SU(3) limit, two octet baryons are  classified by the irreducible representations of the SU(3) group as ${\bf 8}\otimes {\bf 8}  = ({\bf 27}\oplus {\bf 8}_s \oplus {\bf 1})_{S=0} \oplus ({\bf 10}^*\oplus {\bf 10} \oplus {\bf 8}_a)_{S=1} $, where $S$ is the total spin. The system of two $\Lambda$'s  with opposite spins we are interested in belongs to the flavor singlet representation ${\bf 1}$, whose potential is denoted as $V_C^{(1)}$. 
More precisely the single state is denoted as
\begin{equation}
\vert {\rm singlet} \rangle = -\sqrt{\frac{1}{8}}\vert \Lambda\Lambda\rangle + \sqrt{\frac{3}{8}}\vert\Sigma\Sigma\rangle +\sqrt{\frac{4}{8}} \vert N\Xi\rangle .
\end{equation}
Fig.~\ref{fig-6} shows the flavor-singlet potential $V_C^{(1)}(r)$ as a function of $r$ on  volumes $L\simeq$ 2, 3 and 4 fm at $a\simeq 0.12$ fm and $M_{\rm PS} \simeq 1015$ MeV in the SU(3) limit\cite{Inoue:2010es}.  
As can be clearly seen in the figure, the attractive instead of repulsive core indeed appears in this setup.
This result strongly suggests that partial restriction of color combinations due to the Pauli principle among quarks is important for the repulsive core.

\subsection{H dibaryon}
Since the force in the flavor-singlet channel is attractive at all distances, an existence of bound states may be expected. In this channel, a six quark bound state made of $uuddss$, name H-dibaryon, has been predicted\cite{Jaffe:1976yi}, but not observed yet experimentally.
We have investigated whether the H dibaryon appears for the flavor-singlet channel in the SU(3) limit, by using the singlet-potential $V_C^{(1)}(r)$. 
\begin{figure}[hbt]
\centering
\sidecaption
\includegraphics[width=7cm,clip]{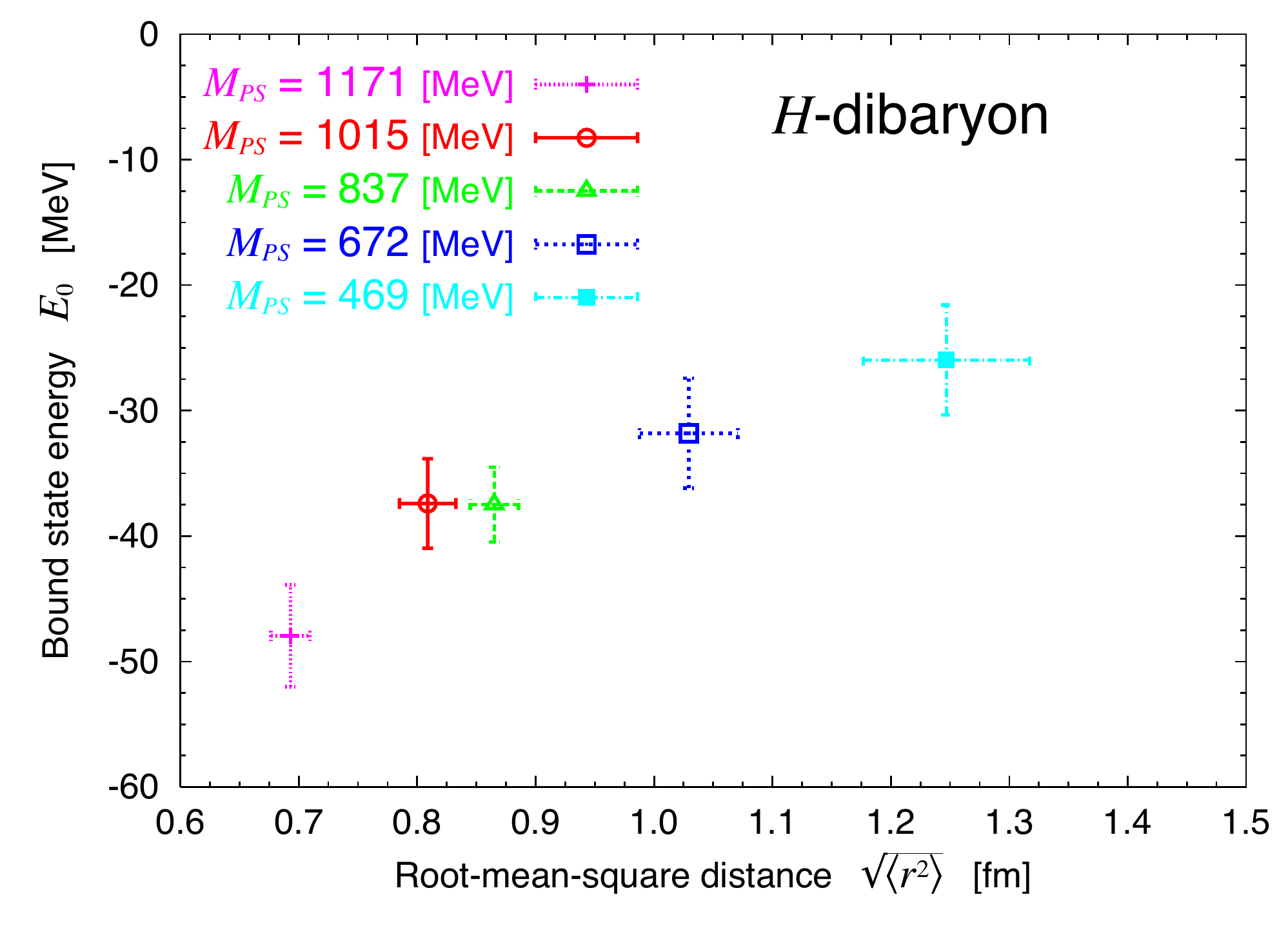}
\caption{The binding energy $B_H=-E_0$ and the root-mean-square distance $\sqrt{\langle r^2\rangle}$ of the bound state in the flavor singlet channel at each $M_{\rm PS}$. Taken from Ref.~\cite{Inoue:2011ai}.  }
\label{fig-7}       
\end{figure}

For this purpose we fit the potential in Fig.~\ref{fig-6} at $L\simeq 4$ fm by the analytic function composed of an attractive Gaussian core plus a long range (Yukawa)$^2$ attraction as $V_C^{(1)}(r)= b_1 e^{-b_2 r^2} + b_3(1-e^{-b_4 r^2})(e^{-b_5 r}/r)^2$. An example of the fitted result with $\chi^2/{\rm dog}\simeq 1$ is shown by the dashed line in Fig.~\ref{fig-6}. By  solving the Schr\"odinger equation with the fitted potential in the infinite volume, we have found one bound state, the H-dibaryon, in the flavor SU(3) limit, whose binding energy and wave function are obtained.
Fig.~\ref{fig-7} shows the binding energy and the root-mean-square(rms) distance of the H-dibaryon obtained from our potential at several values of $M_{\rm PS}$.
The rms distance $\sqrt{\langle r^2\rangle}$ is a measure of the ``size" of the H-dibaryon, which is compared to that of the deuteron, 3.8 fm, in Nature. Although the current result is obtained in the flavor SU(3) limit with heavier up and down  quark masses than Nature, this comparison suggests that H-dibaryon is much more compact than the deuteron.
As $M_{\rm PS}$ decrease, the binding energy decreases and the  rms distance increases.
Note that, despite that the attractive potential becomes stronger as $M_{\rm PS}$ decreases\cite{Inoue:2011ai, Inoue:2010es}, the binding energy $B_H$ decreases du to the fact that the increase of the attraction toward the lighter $M_{\rm PS}$ is compensated by the increase of the kinetic energy for the lighter baryon mass.
As a result, the size of H-dibaryon also increases. 

We close this section by concluding  that the H-dibaryon exists in the flavor SU(3) limit, whose binding energy shows a mild quark mass dependence and is about 20 $\sim$ 50 MeV at this range $M_{\rm PS}$. (See \cite{Beane:2010hg} for the direct calculation of the binding energy for the H-dibaryon in 2+1 flavor QCD.)

\section{Summary and more}
The potential method, called the HAL QCD method, is new but very useful to investigate not only the nuclear force but also general baryonic interaction in (lattice) QCD\cite{Nemura:2008sp,Etminan:2014tya,Yamada:2015cra}.
A comparison of the nuclear potential with the flavor-singlet potential in the flavor SU(3) limit leads to some understanding of the repulsive core, that the Pauli exclusion principle plays an important role for its existence.
The HAL QCD method can be easily extended also to meson-baryon and meson-meson interactions\cite{Ikeda:2013vwa}.  

Our strategy based on the HAL QCD method to investigate nuclear physics and astrophysics is as follows.
We first extract potentials form lattice QCD using the HAL QCD method. 
Using the sophisticated many-body methods combined with these potentials obtained in QCD, 
we then investigate structures of nuclei and nuclear matter, based on which  we finally study phenomena in  astrophysics such as properties of neutron stars and supernova explosions.

As a demonstration how the above strategy indeed works, we show results in Ref.~\cite{Inoue:2013nfe}, which give quark mass dependences of the nuclear matter, the neutron matter and the maximum mass of neutron stars, using nuclear potentials obtained in the flavor SU(3) limit\cite{Inoue:2011ai}.

\begin{figure}[bth]
\centering
\sidecaption
\includegraphics[width=6.5cm,clip]{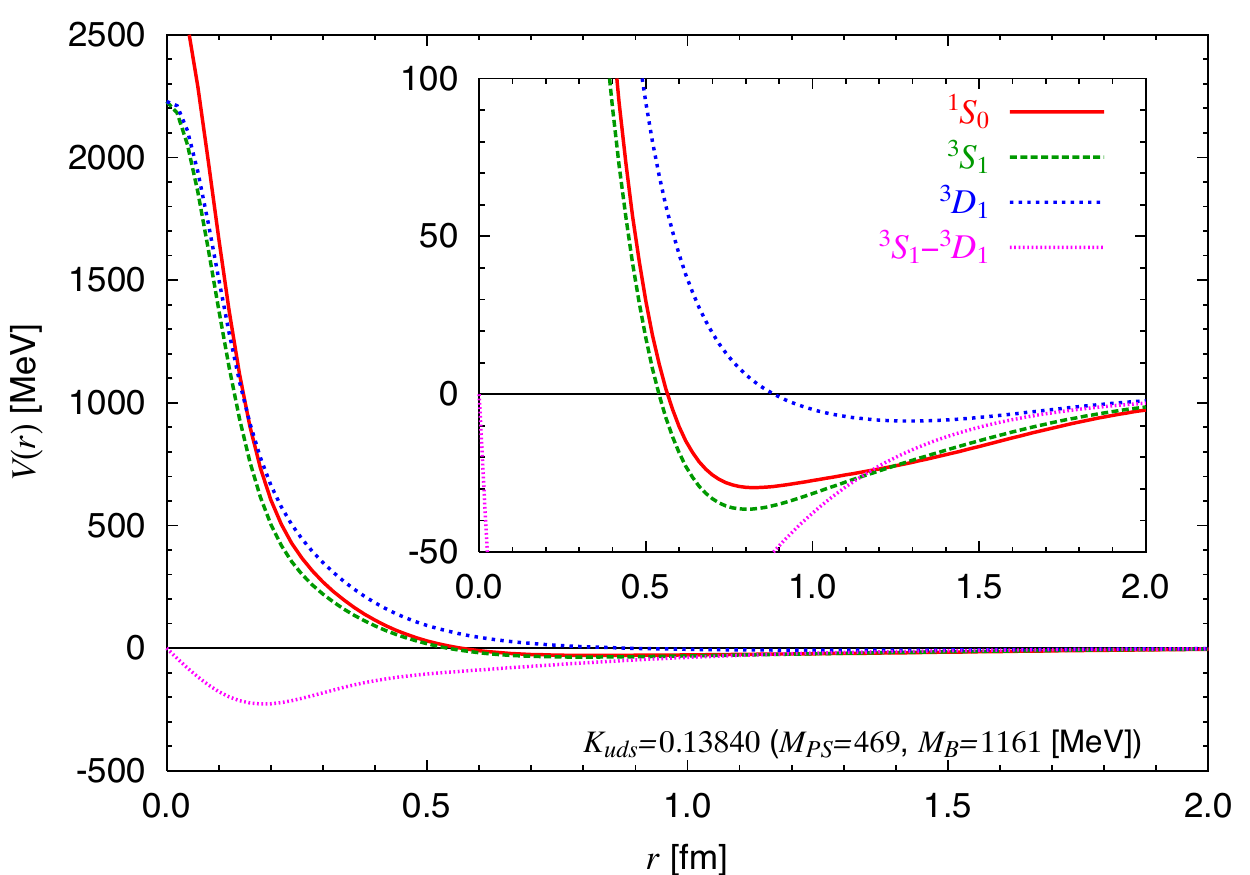}
\caption{The lines represent the $\chi^2$ fits of the nuclear potentials for S and D wave extracted from lattice QCD at $M_{\rm PS} \simeq$ 470 MeV in the flavor SU(3) limit. Taken from Ref.~\cite{Inoue:2013nfe}.  }
\label{fig-8}       
\end{figure}
Fig.~\ref{fig-8} shows the nuclear potentials in S and D waves obtained from fits to the lattice QCD data
in the flavor SU(3) limit at $M_{\rm PS}\simeq 470$ MeV\cite{Inoue:2011ai}. 
\begin{figure}[bth]
\centering
\includegraphics[width=6.0cm,clip]{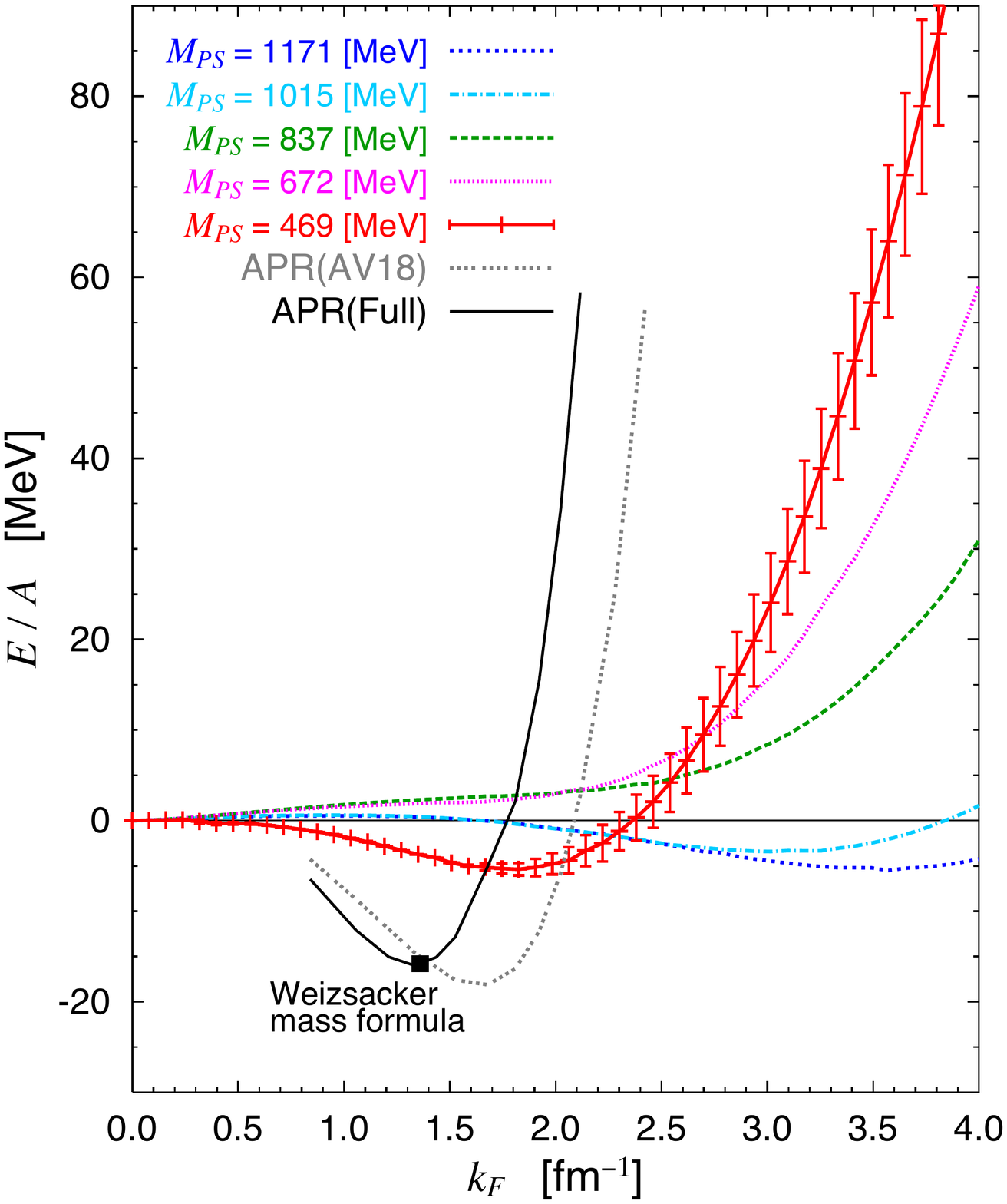}
\includegraphics[width=6.0cm,clip]{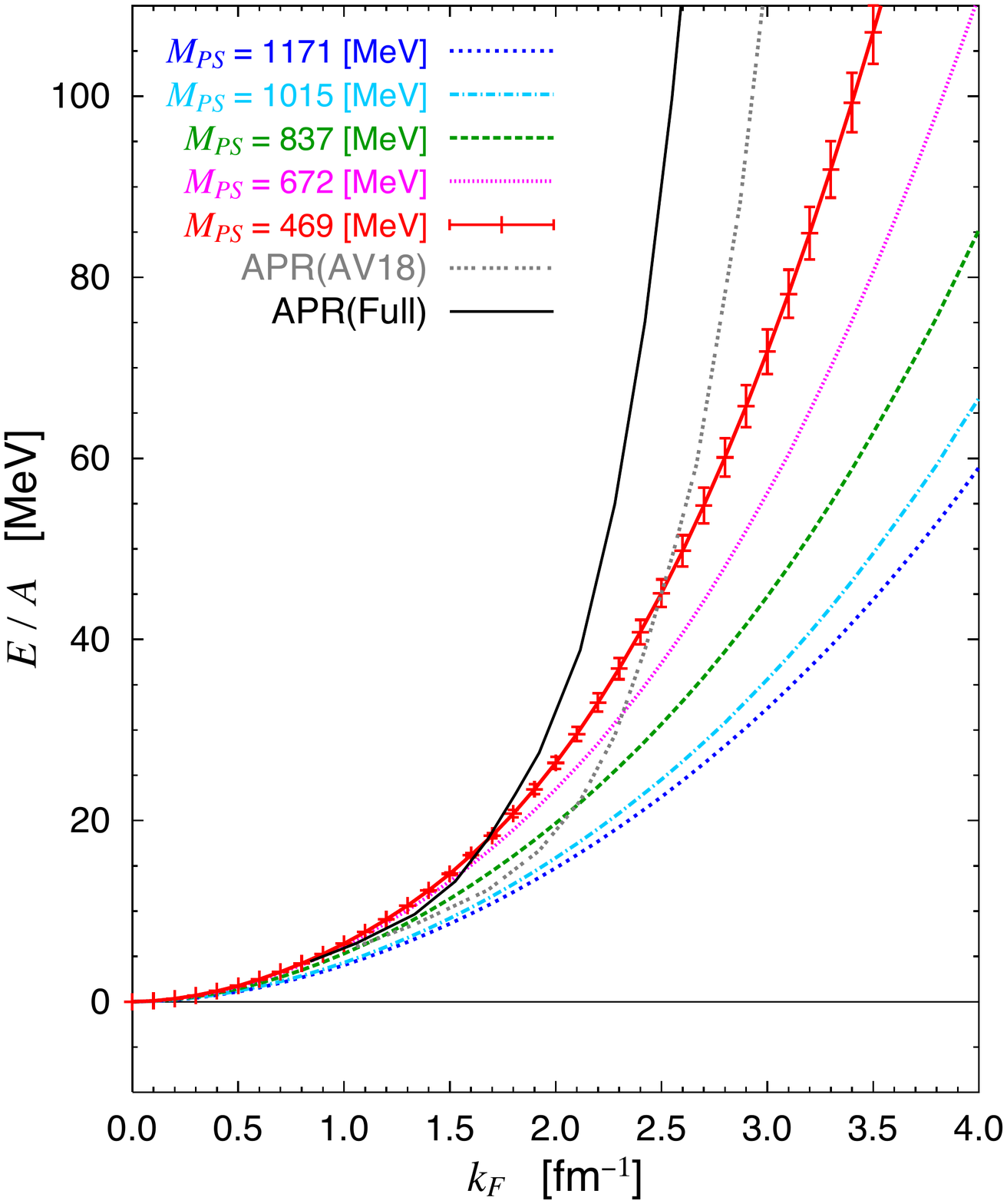}
\vskip -1.0cm
\caption{(Left) Ground state energy per nucleon ($E/A$) for the symmetric nuclear matter as a function of the Fermi momentum $k_F$, obtained by the BHF theory with the nuclear potentials in lattice QCD. Filled square indicates the empirical saturation point, together with the curves labeled as APR from Ref.~\cite{Akmal:1998cf}.  
Errors for the result at $M_{\rm PS}\simeq$ 470 MeV represent statistical uncertainties.
(Right) Ground state energy per nucleon for the pure neutron matter as a function of the Fermi momentum $k_F$. Details of the calculations are the same as the symmetric nuclear matter.
Both are taken from Ref.~\cite{Inoue:2013nfe}.  }
\label{fig-9}       
\end{figure}
Using these potentials, the equation of state (EOS) for the nuclear matter and the neutron matter has been investigated by the Brueckner-Hatree-Fock (BHF) theory\cite{Inoue:2013nfe}.\footnote{Currently, more sophisticated many-body calculations are attempted in collaboration with experts in this area.}
Fig.~\ref{fig-9} (Left) shows the ground state energy per nucleon ($E/A$) for the symmetric nuclear matter ($Z=N=A/2$) with the proton number $Z$, the neutron number $N$ and the mass number $A=N+Z$, as a function of the Fermi momentum $k_F$ for different $M_{\rm PS}$. The most important feature of the symmetric nuclear matter in Nature is its saturation property that $E/A$ has a minimum at normal nuclear matter density $\rho_0$.
The empirical saturation point from the Weizs\"acker mass formula gives $(k_F,E/A) \simeq $ (1.36 fm$^{-1}$, -15.7 MeV),  denoted by the filled square in Figure~\ref{fig-9} (Left), where we also show the results of Ref.~\cite{Akmal:1998cf}, obtained by the variational method using AV18 nuclear potentials\cite{Wiringa:1994wb} with and without phenomenological three-nucleon force.  
Our lattice data show the saturation not only at $M_{\rm PS}\simeq 470$ MeV (the lightest quark mass) but also at $M_{\rm PS}\simeq 1020, 1170$ MeV (the heaviest two quark masses), which indicates that the saturation originates from a suitable balance between the repulsive core and the intermediate attraction of the nuclear force. An appearance of the saturations  at more than $10\rho_0$ at $M_{\rm PS}\simeq 1020, 1170$ MeV,
however, might be an lattice artifact, and needs to be checked.

Fig.~\ref{fig-9} (Right) shows $E/A$ for the pure neutron matter ($N=A$) as a function of $k_F$.
The neutron matter is not self-bound due to large Fermi energy.
The corresponding pressure is calculated as $ P =\rho^2\frac{\partial (E/A)}{\partial \rho}$ with the density $\rho = \frac{k_F^3}{3\pi^2}$, so that  $P$ increases rapidly as $M_{\rm PS}$ decreases.

\begin{figure}[bth]
\centering
\sidecaption
\includegraphics[width=6.5cm,clip]{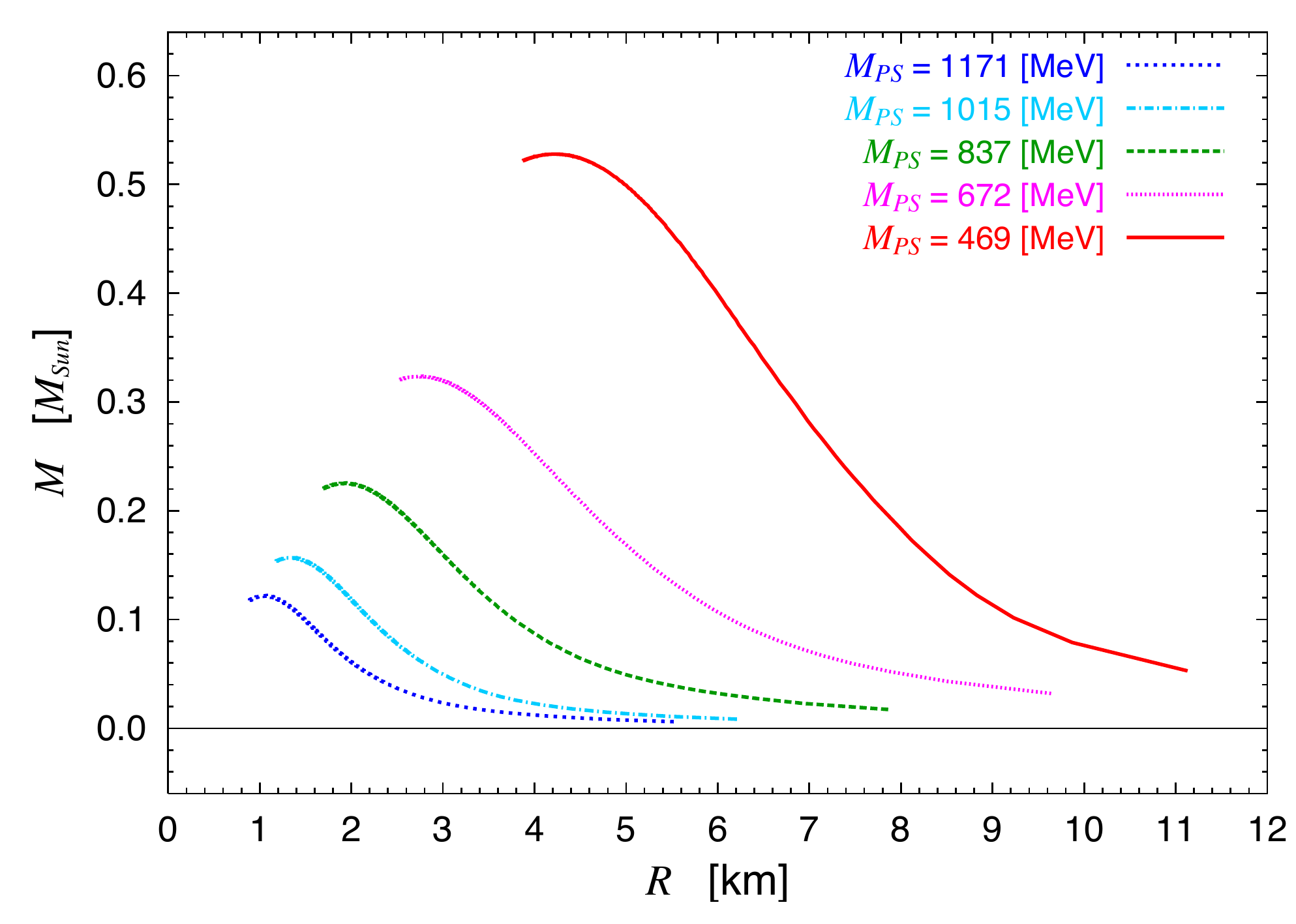}
\caption{Mass-radius relation of the neutron star. Neutron-star matter consists of $n, p, e^{-}$ and $\mu^-$ with charge neutrality and chemical equilibrium. The EOS for the nucleons is obtained by an interpolation between the symmetric nuclear matter and the pure neutron matter under the parabolic approximation. Taken from Ref.~\cite{Inoue:2013nfe}.  }
\label{fig-10}       
\end{figure}
To obtain the mass ($M$) and radius ($R$) relation of the neutron star,
the Tolman-Oppenheimer-Volkoff (TOV) equation is solved 
under the charge neutrality and beta-equilibrium for neutron, proton electron and muon,
with  the EOS of asymmetric nuclear matter approximated as 
\begin{equation}
\frac{E}{A}(\rho,x) =\frac{E_{Z=N}}{A}(\rho) + (1-2x)^2\epsilon_{\rm sym}(\rho), \quad
\epsilon_{\rm sym}(\rho) =\frac{E_{Z=0}(\rho)}{A} - \frac{E_{Z=N}(\rho)}{A}
\end{equation}
for the proton fraction $x=\rho_x/\rho$.

Fig.~\ref{fig-10} shows the $M$-$R$ relation of the neutron star at different $M_{\rm PS}$.
As $M_{\rm PS}$ decreases, the $M$-$R$ curve shifts to the upper right direction, suggesting the stiffening of the EOS.  The maximum mass of the neutron star ($M_{\rm max}$) is found to be 0.53 times the solar mass $M_{\odot}$ at our lightest $M_{\rm PS}\simeq$ 470 MeV, which is too small to account for the observed neutron stars, mainly due to the larger $M_{\rm PS}$. 
A naive extrapolation of $M_{\rm max}$ and the corresponding radius to $M_{\rm PS}\simeq 140$ MeV,
with a function $f(M_{\rm PS})=a/(M_{\rm PS}+b)+c$, 
gives  $M_{\rm max}=2.2 M_{\odot}$ and $R=12 $ km.
To confirm these predictions, we will have to repeat the procedure explained here with the nuclear potentials at $m_\pi \simeq$ 140 MeV.
  
\begin{acknowledgement}
The author would like to thank other members of the HAL QCD Collaboration for discussions.
This work is supported in part by the Grant-in-Aid of the Japanese Ministry of Education (No. 25287046), MEXT SPIRE and JICFuS. 
\end{acknowledgement}

%
%
%
{}

\end{document}